\documentstyle[12pt]{article}

\newcommand{\postscript}[2]{\vspace{5cm}}
\newcommand{\postbb}[3]{\vspace{5cm}}

\begin{document}
\newcommand{\beq}{\begin{equation}}
\newcommand{\eeq}{\end{equation}}
\newcommand{\beqa}{\begin{eqnarray}}
\newcommand{\eeqa}{\end{eqnarray}}
\def\Tr{{\rm Tr}}
\def\lag{Lagrangian}
\def\ra{\rightarrow}
\def\RA{\rightarrow}
\def\x{\times}
\def\etal{{\it etal.,\ }}
\def\sinn{\sin^2 \theta_W}
\newcommand{\RMP}[3]{{\em Rev. Mod. Phys.} {\bf #1}, #2 (19#3)}
\newcommand{\PR}[3]{{\em Phys. Rev.} {\bf #1}, #2 (19#3)}
\newcommand{\PL}[3]{{\em Phys. Lett.} {\bf #1}, #2 (19#3)}
\newcommand{\Rep}[3]{{\em Phys. Rep.} {\bf #1}, #2 (19#3)}
\newcommand{\Ann}[3]{{\em Ann. Rev. Nucl. Sci.} {\bf #1}, #2
(19#3)}
\newcommand{\NS}[3]{{\em Nucl. Sci.} {\bf #1}, #2 (19#3)}
\newcommand{\PRL}[3]{{\em Phys. Rev. Lett.} {\bf #1}, #2 (19#3)}
\newcommand{\NP}[3]{{\em Nucl. Phys.} {\bf #1}, #2 (19#3)}
\newcommand{\con}[3]{{\bf #1}, #2 (19#3)}
\def\mev{\; {\rm MeV} }
\def\MEV{\; {\rm MeV} }
\def\Ev{\; {\rm eV} }
\def\tev{\; {\rm TeV} }
\def\eV{\; {\rm eV} }
\def\gev{\; {\rm GeV} }
\def\etc{ {\it etc}}

\begin{center}

{GRAND UNIFICATION AND THE STANDARD MODEL\footnote{Invited talk presented
at {\it Radiative Corrections: Status and Outlook}, Gatlinburg, TN,
June, 1994.}}
\vspace{3ex}

PAUL LANGACKER  \\ University of Pennsylvania \\ Department of Physics
\\ Philadelphia, Pennsylvania, USA 19104-6396 \\
\today,  UPR-0639T

\vspace{3ex}

ABSTRACT

\end{center}

The current status of gauge unification is surveyed both in the standard
model and its minimal supersymmetric extension.  Implications for proton
decay, Yukawa unification, and the Higgs mass are described.


\section{Introduction}

The precision electroweak data has spectacularly confirmed the electroweak
part of the standard model and has correctly predicted $m_t$ and
$\alpha_s$.  However, the standard model has many shortcomings, most
notably the large number of free parameters, the unexplained fermion
families and masses, and notorious hierarchy and fine-tuning problems.
Because of the many problems it is clear that there must be new physics
beyond the standard model.

There are two general classes of extensions.  One is the possibility of new
layers of compositeness of fermions and/or dynamical symmetry breaking to
replace the elementary Higgs mechanism.  These ideas have run
into significant difficulties which have precluded the construction of
realistic model, and possibly have made the whole line of approach
unlikely.  The problems include the non-observation of large
flavor-changing neutral current (FCNC) effects, which one generally expects
in such models, and the absence of large anomalous contributions to the
$Zb\bar{b}$ vertex, new sources of $SU_2$-breaking (the $\rho_0$ parameter),
or large contributions to the parameters $S_{\rm new}$, $T_{\rm new}$,
$U_{\rm new}$, which describe any new source of physics which contributes
only to the gauge boson self-energies.  In particular, with the probable
discovery of the top quark from CDF and the direct determination of its
mass, it is possible to unambiguously separate the new physics contribution
to these parameters from the effects of $m_t$, with the result that the
limits on $\rho_0$ and on $S_{\rm new}$, $T_{\rm new}$, $U_{\rm new}$ are
stringent.  Another difficulty with compositeness is that one
generally expects new 4-Fermi operators from constituent interchange, and
these have not been observed.

The other generic class of extensions involve some form of unification at a
high scale.  In such schemes one does not expect much new physics at the
TeV scale, except possibly for supersymmetry.  In the viable supersymmetric
models the new particles are usually heavy and decouple both from rare
decays and precision experiments, so that one expects little or no
deviation from the standard model predictions.  In these supersymmetric
extensions one might or might not also expect some form of unification of
the coupling constants at a high scale.

In this talk I will survey the latter set of possibilities,
including the status of the precision electroweak tests and the
determination of parameters, both in the standard model and its
supersymmetric extension.  I will then review some of the implications for
supersymmetric grand unified theories.  It will be seen that the current
data is in generally excellent agreement with coupling constant
unification.  However, the larger value of $m_t$ suggested both by the
recent precision data and the CDF events favors a larger value for $\alpha_s$,
in good agreement with the direct determinations at the $Z$-pole, but
slightly larger than some low energy determinations.  Implications for
proton decay, Yukawa unification, and the Higgs mass will be
briefly discussed.


\section{The Standard Model Parameters}

The high precision electoweak data from LEP and SLC, as well as $M_W$ and
low energy neutral current data, is generally in excellent agreement with
the predictions of the standard model \cite{A}, although there are
discrepancies (at the $ 2 - 3 \sigma$ level) in the SLD measurement of the
left-right polarization asymmetry $A_{LR}$ \cite{B}, and in $R_b = \Gamma
(Z \ra b \bar{b})/ \Gamma (Z \ra {\rm hadrons})$. A global fit
to all indirect data yields \cite{C}
\beqa \sin^2 \hat{\theta}_W (M_Z) &=& 0.2317 (3) (2) \nonumber \\
\alpha_s (M_Z) &=& 0.127 (5) (2) \;\;\;\;\;\;\; (SM) \label{E1} \\
m_t &=& 175 \pm 11^{+17}_{-19} \;\;\;\; {\rm GeV}, \nonumber
\end{eqnarray}
where $\sin^2 \hat{\theta}_W(M_Z)$ and $\alpha_s (M_Z)$ are computed
in the $\overline{MS}$ scheme.  The central values and first uncertainty
are for a Higgs mass of $M_H = 300$~GeV, while the second uncertainty is
from taking $M_H = 1000 \ {\rm GeV} (+)$ and $60 \ {\rm GeV}(-)$.  The
predicted value of $m_t$ in (\ref{E1}) is in remarkable agreement with the
value $174 \pm 16$~GeV suggested by the CDF candidate events \cite{D}.  The
value of $\alpha_s$ is determined from the $Z$ lineshape data.  It is in
good agreement with the value $0.123 \pm 0.006$ obtained from LEP jet event
shapes, but is higher than most low energy determinations \cite{E}, as shown
in Table~\ref{tabI}.  In particular, it is higher than a recent lattice
determination from the $b\bar{b}$ spectrum \cite{G}, which claims a very
small uncertainty.

\begin{table} \centering
\begin{tabular}{|lcc|}  \hline
LEP line-shape    & $0.127 \pm 0.005$   &  \protect\cite{C} \\
LEP event shapes  & $0.123 \pm 0.006$   &  \protect\cite{E} \\
hadronic $\tau$ decays (LEP)  & $0.122 \pm 0.005$  & \protect\cite{E} \\
deep inelastic scattering     & $0.112 \pm 0.005$  & \protect\cite{E} \\
$\Upsilon$, $J/\Psi$ decays   & $0.113 \pm 0.006$  & \protect\cite{E} \\
$c\bar{c}$ spectrum (lattice) & $0.110 \pm 0.006$  & \protect\cite{F} \\
$b\bar{b}$ spectrum (lattice) & $0.115 \pm 0.002$  & \protect\cite{G} \\
\hline
\end{tabular}
\caption[]{Values of $\alpha_s(M_Z)$ extracted from various processes.
Except for the $\tau$ decay value, the low energy determinations
extrapolated to $M_Z$ are lower than the $Z$-pole values.}
\label{tabI}
\end{table}

The lineshape determination of $\alpha_s$ is clean theoretically assuming
the validity of the standard model, but is sensitive to the presence of new
physics which
affects the $Zb\bar{b}$ vertex, as suggested by $R_b$ \cite{C}.  The data
marginally favors a low value for $M_H$, but this is weak statistically
$(M_H < 880$~GeV at 95\% $CL$, or $M_H < 730$~GeV if the CDF $m_t$ value is
included \cite{C}).  Moreover, it is driven almost entirely by the
anomalous values of $A_{LR}$ and $R_b$.  The data also set stringent limits
on the new physics parameters $S_{\rm new}$, $T_{\rm new}$ (or $\rho_0 -1)$
and $U_{\rm new}$ \cite{C}.

In the supersymmetric extension of the standard model (SSM), most of these
conclusions continue to hold.  For most of the allowed parameter space, the
new superpartners and extra Higgs particles are sufficiently heavy that
their contributions to the radiative corrections to the precision
observables are negligible.  That is, one does not expect to see deviations
from the standard model.  However, there is a light Higgs scalar which (for
most of parameter space) acts like a
light standard model Higgs.  One should therefore use the more restricted
range 60~GeV$< M_H < 150$~GeV, rather than 60~--~1000~GeV.  Because of the
strong $m_t - M_H$ correlations in the radiative corrections, one now finds
\beqa \sin^2 \hat{\theta}_W (M_Z) &=& 0.2316 (3) (1) \nonumber     \\
\alpha_s (M_Z) &=& 0.126 (5) (1) \;\;\;\;\;\;\; (SSM) \label{E2} \\
m_t &=& 160^{+11\, +6}_{-12\, -5} \ {\rm GeV} \nonumber \eeqa
rather than (\ref{E1}), where the central value is for $M_H = M_Z$ and the
second uncertainty is for $60$~GeV$<M_H< 150$~GeV.  The $m_t$ prediction is
still consistent with the CDF range, but favors the lower end.


\section{Supersymmetry and Grand Unification}

Let me now turn to supersymmetry and grand unification, which can occur
independently or in combination.  I will first briefly review the ideas of
grand unification and of supersymmetry, and then turn to the possible
unification of coupling constants.

\subsection{Grand Unification}

Some of the shortcomings of the standard model, especially those associated
with the fact that it involves three distinct gauge
sectors, are addressed in grand unified theories~\cite{su5,gutrev}.
The idea is that the
strong, weak, and electromagnetic interactions are unified at some large
unification scale $M_X$, {\it i.e}., the interactions are embedded in a
simple gauge group $G$ which is manifest above this scale.  At
$M_X$ the symmetry is broken to the smaller standard model group $SU_3 \x
SU_2 \x U_1$, so that at low energies the interactions appear different.
If one measures the gauge coupling
constants at low energies and extrapolates to large scales they should meet
at the scale $M_X$ above which the symmetry breaking is irrelevant.  For
unification without gravity to make sense it is necessary that $M_X$ is
small compared to the Planck scale, $M_P \sim G_N^{-1/2} \sim 10^{19} GeV$.

In addition to the gauge interactions, $q,$ $\bar{q}$, $\ell$, and
$\bar{\ell}$ are typically unified, {\it i.e.}, placed in the same
multiplets.  This explains charge quantization (the fact that atoms are
electrically neutral), which is incorporated in, but not explained by, the
standard model.  Moreover, there will be new interactions between quarks
and leptons and their antiparticles which will typically lead to proton
decay.  The simplest example is the Georgi-Glashow model, based on the
gauge group $SU_5$~\cite{su5}.
At a large $M_X$ the symmetry is
broken to $SU_3 \x SU_2 \x U_1$, and then to the unbroken subgroup $SU_3 \x
U_{1Q}$ in a second stage of breaking at the electroweak scale $M_Z$.  The
fermion representations are still quite complicated in this model, with
each generation of fermions placed in a reducible $5^* + 10$.  The $5^*$
representation consists of the left-handed fields $(\nu_e e^- \bar{d})_L$
while the 10 consists of $(e^+ u d \bar{u})_L$.  In addition to the
electroweak and QCD interactions there are two new superheavy gauge bosons
$X$ and $Y$ with electric charges $\frac{4}{3}$ and $\frac{1}{3}$,
respectively, which can mediate transitions between the particles in these
multiplets and lead to proton decay.  For example, the decay $p
\ra e^+ \pi^0$ proceeds by
the exchange of the superheavy $X$ boson, as shown in Figure \ref{fig4}.
The lifetime is
\beq \tau_{p \ra e^+ \pi^0} \sim \frac{M^4_X}{\alpha^2_G m^5_p}, \eeq
where $\alpha_G$ is the value of the coupling constant at the unification
scale.  Around 1980, when such theories were first taken seriously,
the experimental limit on the lifetime was $\geq 10^{30}$~yr.,
corresponding to $M_X > 10^{14}$~GeV.  Several new proton decay experiments
were mounted to search for such decays, but they were not observed,
excluding the simplest $SU_5$ and similar models.
\begin{figure}
\postbb{100 30 500 650}{/u/pgl/fort/nc/graph/misc/proton.ps}{.5}
\caption[]{(a) Diagram for proton decay, $p \ra e^+ \pi^0$, by superheavy
gauge
boson exchange in both ordinary and supersymmetric grand unification. (b)
New dimension-5 mechanism for proton decay emerging in supersymmetric grand
unified theories.}
\label{fig4}
\end{figure}

One can embed the model in larger gauge groups.  In the
$SO_{10}$ model each fermion family is placed in an irreducible 16
dimensional representation, which decomposes under the $SU_5$ subgroup as
$16 = 5^* + 10 +1$.  Each family thus has a new neutrino $\bar{N}_L$, which
is a singlet under the normal electroweak gauge group but which may acquire
a large Majorana mass.  Even larger groups are often considered.  In $E_6$
each family is placed in a 27 dimensional representation, which decomposes
under $SO_{10}$ as $27 = 16 + 10 + 1$.  In addition to the 16 particles
already described, one has a 10 of $SO_{10}$, consisting of a new heavy
down-type quark, $(D \bar{D})_L \sim (D_L D_R)$ for which both left and
right-handed components are $SU_2$ singlets.  The 10 also contains a
vector doublet of leptons $(E^+ E^0 \bar{E}^0 E^-)_L \sim (E^+E^0)_L
(E^+E^0)_R$, for which both the left and right components transform as weak
doublets.  There is also a second neutral lepton, $S_L$, which again may
acquire a large Majorana mass.

\subsection{Supersymmetry}

Supersymmetry~\cite{susy}
is a new type of symmetry relating fermions to bosons, which
can occur with or without grand unification.  If unbroken, it would mean
that for every fermion there is a boson with the same mass and related
couplings.  In the real world the symmetry is broken.

Although supersymmetry has not been observed experimentally, there are
several motivations for believing that it might exist.  The first is that
it can stabilize the weak scale.  In the absence of supersymmetry there are
quadratically divergent loop corrections to the mass of the Higgs which
tend to renormalize the mass up to very large scales such as the Planck
scale.  One has to do a fine-tuned cancellation of those corrections
against the bare terms to keep the observed weak interaction
scale.  In the presence of supersymmetry, however, there will be
cancellations between loops involving fermions and bosons.  The
cancellation would be exact if the supersymmetry were unbroken.  In the
presence of breaking there is still a sufficiently stable weak scale
provided $M_{\rm SUSY} <{\rm O} (1$~TeV), where $M_{\rm SUSY}$ is a typical
mass of the new particles.

The second motivation is that in gauged supersymmetry, known as
supergravity, there is an automatic unification of the other interactions
with gravity.  This does not by itself make gravity renormalizable, but at
least brings it into the game.  A final motivation is that the observed
coupling constants are consistent with the simplest version of
supersymmetric grand unification.

There are many consequences of supersymmetry.  One is that there must be a
second Higgs doublet, so the spectrum of the theory will involve additional
charged and neutral Higgs particles.  However, there is always one light
Higgs scalar satisfying
\beq M^2_{H^0} < \cos^2 2 \beta \;\; M^2_Z + {\rm H.O.T.} \;\; (O(m_t^4))<
(150 \;{\rm GeV})^2 \label{eq2},\eeq
where $\tan \beta = v_t/v_b$ is the ratio of vacuum expectation values of
the two Higgs doublets which generate masses for the $t$ and $b$,
respectively.  This is in contrast to the standard model, in which there is
no rigorous upper-bound on the Higgs mass, though there are reasonably
convincing theoretical arguments that suggest $M_{H^0} < 600 - 1000\; {\rm
GeV}$~\cite{higgs}.  The first term
in (\ref{eq2}) appears at tree level in the minimal
supersymmetric extension of the standard model~\cite{susyhiggs}
(slightly higher values are
allowed in non-minimal models with additional
Higgs singlets~\cite{singlet}).  If this
were the only term it would bound the light Higgs scalar mass to be less
than $M_Z$.  However, there are large loop corrections~\cite{susyhiggs}
which can scale like
$m_t^4$, so that one typically has an upper limit around 130 (150)~GeV in
the minimal (non-minimal) model.  For most of parameter space the second
Higgs doublet is heavier.

In addition, there are the new superpartners for every known particle:  for
each quark there must be a scalar quark, $\tilde{q}$, and each lepton must
be associated with a new scalar lepton, $\tilde{\ell}$.  Similarly, the
gauge bosons have new fermionic partners; for example, the $W$ has a wino
partner, $\tilde{w}$.  Typically, these new particles will be in the
several hundred GeV range.  It is often the case that there is a lightest
supersymmetric partner (LSP), and in many versions of the model this would
be an excellent candidate for cold dark matter~\cite{cdm}.
It is interesting that the
most plausible mechanism for breaking electroweak symmetry in
supersymmetric models requires a large $m_t$, comparable to the values
needed by the precision experiments and by CDF.


\subsection{Gauge Unification and Its Implication}
\subsubsection{Unification of Coupling Constants}

Using the observed low energy gauge couplings as boundary conditions one
can compute the running couplings from the renormalization group equations
\beq \frac{d\alpha_i^{-1}}{d \ln \mu} = - \frac{b_i}{2\pi} - \sum^3_{j=1}
\frac{b_{ij} \alpha_j}{8\pi^2}, \label{eq3} \eeq
where $\alpha_i = g_i^2/4\pi$ and $g_i$ is the gauge coupling of the
$i^{\rm th}$ gauge group.  The two terms on the right are
respectively the one and two-loop contributions to the running.
Equation (\ref{eq3}) may be solved analytically to yield
\beqa \alpha_i^{-1} (\mu) &=& \alpha_i^{-1} (M_X) - \frac{b_i}{2\pi} \ln
\left( \frac{\mu}{M_X} \right) \nonumber \\
&& + \sum^3_{j = 1} \frac{b_{ij}}{4\pi b_j} \ln \left[ \frac{ \alpha_j^{-1}
(\mu)}{\alpha_j^{-1} (M_X)} \right] . \eeqa
For the observed couplings the first two terms dominate and the inverse
coupling runs approximately linearly in $\ln \mu$.  However, the nonlinear
correction from the last (two loop) term is not completely negligible.

If
there is a grand unification then naively the three gauge couplings should
have a common value at the unification scale $M_X$.  More precisely, one
expects
\beq \alpha_i^{-1} (M_X) = \alpha_G^{-1} (M_X) + \delta_i + \Delta_i, \eeq
where the first term represents the common coupling and the other two are
threshold corrections~\cite{thresh}.
$\delta_i$ represents low-scale corrections due to
the fact that not all of the new particles are degenerate.  There are
contributions from $m_t > M_Z$, as well as from the nondegeneracy of the
new superpartner and heavy Higgs particles, $M_i^{\rm new} \neq M_{\rm
SUSY}$.  The last term represents the high-scale threshold corrections, due
both to mass splittings at the high scale, $M_{\rm heavy} \neq M_X$, and to
nonrenormalizable operators (NRO) which may be left over from quantum
gravity.

The coefficient functions depend on the matter content of the theory.  If
one has only the standard model particles above the $Z$ scale, then the
one-loop coefficients are~\cite{coef}
\beq b_i = \left( \begin{array}{c} 0 \\ -\frac{22}{3} \\ -11 \end{array}
\right) \; + \; F \left( \begin{array}{c} 4/3 \\ 4/3 \\ 4/3 \end{array}
\right) \; + N_H \left( \begin{array}{c} 1/10 \\ 1/6 \\ 0 \end{array}
\right).\eeq
The three terms refer, respectively, to the contributions of gauge boson,
fermion, and Higgs loops;  $F$ is the number of fermion generations; and
$N_H$ is the number of Higgs doublets.  Similarly, in the
minimal supersymmetric extension (MSSM) the new superpartners and Higgs
bosons modify the equations so that
\beq b_i = \left( \begin{array}{c} 0 \\ -6 \\ -9 \end{array}
\right) \; + \; F \left( \begin{array}{c} 2 \\ 2 \\ 2 \end{array}
\right) \; + N_H \left( \begin{array}{c} 3/10 \\ 1/2 \\ 0 \end{array}
\right).\eeq
The two-loop coefficients, $b_{ij}$ may be found in \cite{coef}.

To apply the renormalization group equations one uses the observed
couplings at the electroweak scale as boundary conditions.  The
$SU_3 \x SU_2 \x U_1$ couplings are given by
\beq g_3 = g_{s}\;\; ,\;\; g_2 = g \;\; , \;\; g_1 = \sqrt{\frac{5}{3}} g',
\eeq
where the coefficient in $g_1$ is a normalization factor, needed so that
all of the charges which unify are normalized in the same way.  That is,
\beq \Tr  Q^2_s  = \Tr  Q^2_2  = \frac{5}{3} \Tr
\left(\frac{Y}{2}\right)^2.\eeq
Since historically the weak hypercharge $Y$ was renormalized differently,
this must be compensated for.  The observed charge of the positron and weak
angle are related by
\beq e = g \sin \theta_W \eeq
and
\beq \sin^2 \theta_W = \frac{g'^2}{g^2 + g'^2} = \frac{g_1^2}{ \frac{5}{3}
g^2_2 + g_1^2}.\eeq
At the unification scale one has $g_1 = g_2$, implying $\sin^2\theta_W \ra
\frac{3}{8}$~\cite{gqw}.
At low energies the couplings will be different, yielding a
smaller value for $\sin^2\theta_W$. Hence,
\beq \alpha_3 = \alpha_s  , \ \ \ \; \alpha_2 = \frac{\alpha}{\sin^2 \theta_W}
\; , \ \ \ \; \alpha_1 = \frac{5}{3} \; \frac{\alpha}{\cos^2 \theta_W} \eeq
where all quantities are to be evaluated at $M_Z$ in the $\overline{MS}$
scheme.  I will use as input values~\cite{fks}
\beq \alpha(M_Z)^{-1} = 127.9 \pm 0.1, \eeq
where the uncertainty  is due
to low energy hadronic uncertainties, and~\cite{C}
\beq \sin^2 \hat{\theta}_W (M_Z) = 0.2316 \pm 0.0003, \eeq
as determined from the precision data.  I will usually use these to
predict $\alpha_s$, which is the least known.  However, for those tests in
which $\alpha_s$ is an input I will use the range
\beq \alpha_s (M_Z) = 0.12 \pm 0.01 \eeq
which is a reasonable average of both the LEP  and low
energy determinations.

{}From these inputs one can plot the running couplings and see whether they
meet.  It is seen in Figure~\ref{fig1} that they do not meet in the
ordinary standard model, but they do meet with reasonable precision in the
supersymmetric extension~\cite{susycoup}-\cite{couplnew}.
This provides evidence, independent of the
nonobservation proton decay, that the simplest versions of nonsupersymmetric
grand unification are excluded, while the supersymmetric case is allowed.

\begin{figure}
\postbb{50 70 530 680}{/u/pgl/fort/nc/graph/gut/xxgut.ps}{.6}  
\caption[]{Running coupling constants in the standard model and in the
minimal supersymmetric extension. From~\cite{pol1}.}
\label{fig1}
\end{figure}

\subsubsection{Implications of Coupling Unification}

The coupling constants unify in the supersymmetric extension of
the standard model at a scale
\beq M_X \sim 3 \x 10^{16} {\rm GeV}, \eeq
but not in the ordinary standard model.  Does this constitute proof of
supersymmetry?  No, of course not.  The apparent unification could be an
accident, having nothing to do with grand unification.  Similarly,
ordinary models without supersymmetry could be modified to force
unification, for example, by introducing intermediate scales or new
multiplets of particles split into light and heavy sectors.  Of course,
such models are ad hoc and have no predictive power for the gauge
couplings.  Such new ingredients could also disturb the agreement within
the supersymmetric extension.

The predictions are largely independent of the specific GUT group, $SU_5$,
$SO_{10}$, $E_6$ etc., provided the charge normalizations are
preserved.  Also, the predictions do not depend on the number of families
at 1-loop order, although weak dependence does enter at 2-loops.  This is
because a family is a complete multiplet of the underlying GUT, which
affects the slopes of all of the gauge couplings equally.  In contrast,
there is a strong dependence on the number of Higgs doublets, because the
latter involves multiplets splits into light and superheavy components.
Only the light components (the Higgs doublets) affect the running, and
these contribute to the running of the electroweak but not the strong
couplings.  One could improve the unification of couplings in the
ordinary grand unified theory by adding additional Higgs doublets, but
only at the expense of having a lower unification scale and therefore a
more rapid (and unacceptable) proton decay rate.

So far, the predictions are idealized and ignore the threshold corrections.
A number of authors~\cite{robross}-\cite{thunc}
have pointed out that there are irreducible theoretical
uncertainties in the predictions.  One source is from the low scale
thresholds.  It was emphasized by Roberts and Ross~\cite{robross}
that the approximation
made in the early papers of treating all of the new particles as degenerate
with a common mass $M_{\rm SUSY}$ is not adequate, and that
the splittings between the sparticles
are more important.  For example, the colored
particles are usually heavier than the uncolored ones, and this is
more important than the average mass because it discriminates between the
gauge couplings.  There are also uncertainties associated with the
splittings of the superheavy particles about the unification
scale~\cite{pol1,thunc}.
Also, since the unification scale is two or three orders of magnitude below
the Planck scale there may be significant nonrenormalizable operators
(NRO), typically of $O(M_X/M_P)\sim 10^{-1} - 10^{-3},$ which may
contribute to the relative normalization of the gauge
couplings~\cite{nro,pol1,thunc}.

It is not convenient to display these uncertainties on the plot of the
three couplings.  It is more useful to use two of the couplings to predict
the third, allowing one to show the theoretical uncertainties explicitly.
Traditionally, people have used $\alpha$ and $\alpha_s$ to predict $\sin^2
\theta_W$.  However, that is not the optimum approach because the largest
uncertainties are then in the input quantity, $\alpha_s$.  Furthermore,
different authors have used different values of $\alpha_s$, leading to
considerable confusion.
It is more enlightening to use
$\alpha$ and $\sin^2\theta_W$ to predict $\alpha_s$, leading to the
prediction shown in Figure~\ref{fig2}.
\begin{figure}
\postbb{65 75 525 680}{/home/pgl/fort/nc/graph/gut/xxagut.ps}{0.6} 
\caption[]{ Predictions for $\alpha_s$ using $\alpha$ and $\sin^2\theta_W$ as
input quantities in both the standard model and in the MSSM. The large
theoretical error bar includes a reasonable (but not rigorous) estimate of
the low and high scale threshold uncertainties
and of NRO's.  The smaller error bars
assume that the new particles are all degenerate at $M_Z$ or 1~TeV and
ignore the high scale uncertainties. From~\cite{pol1}.}
\label{fig2}
\end{figure}
The standard model prediction~\cite{pol1}
\beq \alpha_s = 0.073 \pm 0.001 \pm 0.001 \label{eq8b} \eeq
is far from the experimental data.  In (\ref{eq8b}) the first uncertainty
is from the inputs and $m_t$, while   the second is from the high-scale
thresholds and NRO.  The prediction in the supersymmetric
extension is~\cite{pol1}
\beq \alpha_s = 0.129 \pm 0.002 \pm 0.005^{+0.005}_{-0.002} \pm 0.006,
\label{18}\eeq
where the uncertainties refer respectively to the inputs at
$m_t$; the low scale (SUSY) thresholds; the high scale thresholds; and NRO.
It must be emphasized that the uncertainties are only estimates, based on
reasonable ranges for the magnitudes of the mass splittings and NRO
coefficients.  They should be interpreted as typical values, not absolute
error bars.  The central value in (\ref{18}) is in good agreement with the
higher values of $\alpha_s$ determined from the $Z$ line shape and JET
event topologies but is somewhat higher than some of the low energy
determinations.  If the latter turn out to be true then supersymmetric
unification would require large but not unreasonable threshold corrections.
In Figure~\ref{fig2} the predicted point includes the full uncertainties,
combined in quadrature.  The theoretical uncertainties would be much
smaller if one ignored the high scale threshold and NRO's and assumed that
all of the new particles are degenerate with a mass either $M_Z$ or 1~TeV.
However, the larger error bar indicated in the figure is much more
reasonable.  One can also adopt the more traditional approach of using $\alpha$
and $\alpha_s$ to predict $\sin^2 \theta_W$ as shown in
Figure~\ref{fig3}.

\begin{figure}
\postbb{60 220 530 690}{/u/pgl/fort/nc/graph/gut/xxszgut.ps}{.6}
\caption[]{Prediction of $\sin^2\theta_W$ using $\alpha$ and $\alpha_s$ as
inputs, both in the standard model and in the MSSM. From~\cite{pol1}.}
\label{fig3}
\end{figure}

If the apparent coupling unification is not just an accident, then,
barring fortuitous cancellations, there
are very few types of new physics other than supersymmetry which would
be allowed and not mess up the unification.
These include new gauge
structures which commute with the standard model, such as new heavy $Z'$
bosons; new complete multiplets of fermions and their superpartners,
including sequential, mirror, or exotic families; and new gauge singlets.

The connection of all of this with superstring theories is somewhat
obscure.  The basic ideas of supersymmetric coupling unification are
consistent with the ideas of superstrings.  However, the unification scale
$M_X \sim 3 \x 10^{16}$~GeV is smaller then what one would expect in a
naive string theory, where typically $M_X^{\rm string} \sim g \x 5 \x
10^{17}$~GeV~\cite{string}.
On the one hand, the fact that $M_X$ is lower than the
gravity scale implies that the unification of the microscopic interactions
without gravity is consistent.  It is not at all clear whether the observed
pattern can emerge consistently from a superstring theory.  It is
conceivable that it does and that there are large string-scale threshold
corrections~\cite{stringthresh},
but so far no realistic explicit models.  It is hard, but not
impossible, to imagine a real grand unified theory emerging below the
string scale, but so far nobody has found a successful compactification.
Attempts have yielded a large number of unwanted new particles and no
mechanism for breaking the grand unification symmetry.  For these reasons I
will concentrate on true supersymmetric grand unified theories, and the
connections with the more elegant possibility of string theories must wait.

There are a number of implications, predictions, and problems for a true
grand unified theory.  One possible difficulty (which is not necessarily
shared by superstring theories) is proton decay.  Proton decay is
especially problematic for ordinary (nonsupersymmetric) grand unified
theories~\cite{gutrev},
where the $SU_3$ and $U_1$ couplings unify at the relatively low
scale $M_X \sim (2 - 7) \x 10^{14}$~GeV.  (The $\sin^2 \theta_W$ prediction
does not work in such theories.)  From diagrams such as shown in
Figure~\ref{fig4}a the predicted lifetime is of order $\tau_{p \ra e^+
\pi^0} \sim 10^{31 \pm 1}$, which is excluded by the experimental limit
$\tau_{p \ra e^+ \pi^0} > 10^{33}$~yr~\cite{taupro}.

In the supersymmetric grand unified theories the unification scale is much
higher, typically $M_X \sim (1 - 5) \x 10^{16}$~GeV.  This leads to a
safely unobservable decay rate by the ordinary superheavy boson exchange,
$\tau_{p \ra e^+ \pi^0}\sim 3 \x 10^{38 \pm 1}$~yr.  However, the
supersymmetric models have a new mechanism, the dimension-5 operators
mediated by (fermionic) superheavy higgsinos,
as shown in Figure~\ref{fig4}b.
Since the exchanged heavy particle is a fermion one has a much more rapid
decay with lifetime $\tau_p \sim M_X^2$, implying $\tau_{p \ra \bar{\nu}
K^+}\sim 10^{29 \pm 4}$~yr~\cite{susylife}.
Much of this range is excluded by the
experimental limit $\tau_{p \ra \bar{\nu} K^+ } > 10^{32}$~yr.  Many
specific models are either pressed or eliminated.  Again, I stress that
such mechanisms may not be present in superstring theories, which do not
have the full structure of grand unification.

Many of the simpler grand unified theories (those in which the fermion
masses are generated by Higgs bosons in the fundamental representation)
predict Yukawa unification between the $b$ and $\tau$
couplings~\cite{yukawa}, {\it
i.e.}, $m_b (M_X) = m_\tau (M_X)$ at the unification scale.  At low
energies, due to different renormalizations of the $b$ and $\tau$ vertices,
one has $m_b \sim (2 - 3) m_\tau$.  A number of authors have studied
whether this actually holds~\cite{bp}-\cite{yukstud}.
The difficulty is that the predicted $m_b$
tends to be large due to gluonic vertex corrections, which are large if
$\alpha_s$ is.  They are partially cancelled by Higgs mediated corrections
involving the top quark Yukawa.  We have seen that the $\alpha_s$ predicted
by grand unification is on the high side of the data, and this in turn
implies relatively large predictions for $m_b$.  Consistency suggests that
one should use the prediction for $\alpha_s$ from gauge unification rather
than a fixed input value.\footnote{Threshold corrections affect both
$\alpha_s$ and the $m_b$ prediction, and should be treated
consistently~\cite{pol2}.}
The result is that
two narrow bands in the ratio $\tan \beta \equiv v_t/v_b$ of the
two Higgs vacuum expectation value are consistent with the data, as shown
in Figure~\ref{figu6}.
\begin{figure}
\postscript{/u/pgl/fort/nc/graph/misc/Fmttgb.ps}{.5} 
\caption[]{Regions in the $\tan \beta -  m_t$ plane allowed by Yukawa
unification. $m_b$ is too large in between the small and large $\tan \beta$
branches while outside one of the Yukawa couplings diverges below the
unification scale.  Also shown is the band predicted by the unification of
all three Yukawa couplings, which occurs in some specific models.
{}From~\cite{pol2}.}
\label{figu6}
\end{figure}

The simplest supersymmetric grand unified theories predict Yukawa
unification, and are only viable for two bands in $\tan \beta$.  In
particular, one of the solutions implies that $\tan \beta$ is close to
unity.  These predictions need not hold in superstring-inspired imitators,
but only in true grand unified theories.  However, the same
prediction for the Yukawa couplings fails for the first two families, and
in practice one must invoke some sort of perturbation~\cite{gutrev}
on the theory to
correct the masses of the first two families, which presumably would not be
large for the third family.  Some specific models, in particular $SO_{10}$
models in which the fermion masses are generated by a single complex
ten-dimensional Higgs multiplet, make the stronger prediction of three
Yukawa unification, $m_b = m_\tau = m_t$ at the Planck scale~\cite{yuk3}.
This is
consistent with the observed value of $m_t$ only for the large $\tan \beta$
branch.  This is a much more model dependent assumption.

Supersymmetric models may have strong implications for the mass of the
lightest Higgs scalars.  Almost all supersymmetric models imply the upper
bound $M_{H^0} < 150$~GeV on the lightest Higgs scalar.  However, the
models with Yukawa unification on the small $\tan \beta$
branch have a stronger prediction~\cite{bp2}-\cite{pol3}.
For $\tan \beta \sim 1$ the tree-level
contribution to the mass approximately vanishes, so that the masses are
generated mainly by the loop corrections.  One finds the more stringent
upper limit~\cite{pol3}
\beq M_{H^0} < 110 \;{\rm GeV}.\eeq
Although there are many assumptions, namely supersymmetric unification,
Yukawa unification, and the small $\tan \beta$ branch, this is a striking
prediction. If true, it may be possible to observe the Higgs at LEP-II or
possibly even LEP-I.  For most allowed value of the supersymmetry
breaking parameters the mass is much less than the upper limit, as is shown
in Figure~\ref{fig10}.  The various limits on the Higgs mass as a function
of $m_t$ are shown in Figure~\ref{fig11}.
\begin{figure}
\postscript{/u/pgl/fort/nc/graph/misc/Fc5fig06a.ps}{.6}
\postscript{/u/pgl/fort/nc/graph/misc/Fc5fig06b.ps}{.6}
\postscript{/u/pgl/fort/nc/graph/misc/Fc5fig06c.ps}{.6}
\caption[]{Predictions for the lightest Higgs mass for various ranges of
$m_t$ as the supersymmetry breaking parameters  are chosen
randomly.  The Higgs mass is typically well below the upper limit of
110~GeV. From~\cite{pol3}.}
\label{fig10}
\end{figure}
\begin{figure}
\postscript{/u/pgl/fort/nc/graph/misc/Fupperbound.ps}{.6}
\caption[]{Generic upper limits of 130 and 150~GeV in the minimal
supersymmetric standard model (MSSM)~\cite{susyhiggs}
and in  extended models (NMSSM)
involving additional Higgs singlets~\cite{singlet}.
Also shown are the more detailed
upper limits on the lightest Higgs mass  as a function of $m_t$ in the MSSM
 and in the constrained models respecting Yukawa unification with
 $\tan \beta$ close to 1.  Also shown is  the {\em lower} bound on the standard
model Higgs boson mass implied  by vacuum
stability~\cite{sher,arnold}.
(In the standard model, large Yukawa couplings associated with
large $m_t$ tend to destabilize the vacuum.)}
\label{fig11}
\end{figure}

{}From Figure \ref{fig11} one sees that there are upper bounds on the Higgs
mass in the supersymmetric models, and there is also a lower bound from
vacuum stability in the standard model~\cite{sher,arnold}.
There is actually a forbidden gap,
at least in the restricted model with Yukawa unification.

Another implication of grand unification is that typically one generates
small masses for the ordinary neutrinos by a seesaw mechanism~\cite{seesaw}.
If the
large scale in the seesaw is a typical grand unification mass, or a few
order of magnitude smaller, then one expects masses that are in the range
relevant to solar neutrino oscillations and perhaps a component of hot dark
matter associated with $m_{\nu_\tau}$~\cite{bkl}.
This connection is very exciting,
but the details are extremely model dependent~\cite{cl,bs}.

Another aspect of supersymmetric theories is that typically the
superpartners and additional Higgs particles have masses much greater than
$M_Z$.  In that case they decouple from precision experiments and one does
not expect to see large deviations from the standard model predictions.

The only successful scheme for supersymmetry breaking is the hidden sector
scenario~\cite{susy},
in which the supersymmetry is broken in a hidden sector which has
no interactions with the ordinary particles except gravity.  The breaking
is transmitted to the observable world only by very weak gravitational
interactions.  Typically, this implies that all of the scalars in the
theory will have a common mass evaluated at $M_p / \sqrt{8 \pi}$, and that
the gauginos will also have a common mass at that scale.  At lower energies
the masses diverge due to running effects, and if $m_t$ is sufficiently
large one of the Higgs masses will be driven negative at low energies,
leading to radiative $SU_2 \x U_1$ breaking~\cite{susy}.
This elegant connection
between the supersymmetry and $SU_2 \x U_1$ breaking requires a large
$m_t$.  The fact that the scalars start out degenerate at a
high scale means that flavor changing neutral current (FCNC) effects at low
energy are relatively small.  The latter are generated by box diagrams
involving the superpartners.  They are proportional to mass splittings and
are sufficiently suppressed in the hidden sector models, at least for
transitions involving the first two families.
There are several new sources of CP violation in supersymmetric models, in
addition to the usual CKM phases~\cite{cp}.
In the hidden sector scenario the
neutron electric dipole moment $d_n < 10^{-24} e-$cm places severe
constraints on the phases of the supersymmetry breaking terms.  The most
plausible solution is that the supersymmetry breaking parameters are
real, which is not very surprising if they are generated by gravity.
However, if they are real there will usually be no observable new sources
of CP violation in the MSSM.

Although the hidden sector models present a nice
general picture, the detailed mechanism for the supersymmetry breaking in
the hidden sector is unclear.

There are many implications for the sparticles and the second Higgs
doublet, but the predictions for their
spectra are model dependent~\cite{spectrum}.  In
most such models there is a lightest supersymmetric particle (LSP).  If
this is neutral it will be a candidate for cold dark matter~\cite{cdm}.
Recently, it
has been emphasized that the usual assumption that the hidden sector models
imply a common scalar mass at the unification scale $M_X$ will not
necessarily hold~\cite{pp}.
It might be more plausible to assume that the common
scalar mass is at $M_p/\sqrt{8 \pi}$.  In that case, the running between
$M_p/\sqrt{8\pi}$ and $M_X$ can have a nontrivial effect
on the low energy theory.

One of the most important problems in the standard model is an explanation
of the fermion masses and mixing.  Unfortunately, neither ordinary nor
supersymmetric grand unified theories yield much information on this.  (The
one thing that they do predict is the ratio $m_b/m_\tau$ in those models
with Yukawa unification.)  If there is really an underlying superstring
theory then the fermion mass spectrum will ultimately be
determined by the compactification of the extra dimensions.
However, nobody has concrete and realistic models.  Recently, there has been
considerable activity in texture models, in which one postulates a form of
the fermion mass matrices at high energies and then makes predictions for
low energies~\cite{texture}.  Much of this work has been quite successful
phenomenologically.  However, the form of the initial textures of the mass
matrices are model dependent, and to implement them requires the addition
of extra symmetries.  Usually, they require higher-dimension Higgs
multiplets, which are not compatible with the simplest superstring
theories.


\section{Conclusions}

\begin{itemize} \item The precision electroweak experiments have confirmed
the electroweak standard model spectacularly even though there are $2
- 3$ $\sigma$ deviations in $R_b$, and $A_{LR}$.

\item Currently the weak angle is determined to be~\cite{C}
\beqa \overline{MS} & \; & \hat{s}^2_Z = 0.2317 (3)(2) \nonumber \\
{\rm on-shell} && s_W^2 = 0.2243 (12) = 1 - \frac{M_W^2}{M_{Z}^2} \\
{\rm effective} && \bar{s}_{\ell}^2 = 0.2320(3)(2) = \kappa_\ell
\hat{s}^2_Z \nonumber \eeqa
in various renormalization schemes, where the central value and first
uncertainty are for $M_H = 300$~GeV, and the second uncertainty is for 60
GeV $< M_H < 1000$~GeV.

\item One predicts
\beq m_t = 175 \pm 11^{+17}_{-19} {\rm GeV} \eeq
in the standard model, where the second uncertainty is from $M_H$.  This is
in spectacular agreement with the CDF direct candidate event masses $174
\pm 16$~GeV.

\item The predictions change slightly
in the supersymmetric extension due to the fact that the Higgs mass is
lower.  There one predicts
\beq m_t = 160^{+11+6}_{-12-5} \;{\rm GeV},\eeq
which is on the lower side of, but still consistent with, the CDF range.
Also, $\hat{s}^2_Z \ra 0.2316 (3) (1)$.

\item  One can also determine the value of the strong coupling at the
$Z$-pole, namely
\beq \alpha_s (M_Z) = 0.127 (5)(2) \label{eqd}\eeq
from the $Z$ line shape.  This is in good agreement with the value obtained
from jet event shapes at LEP, $0.123 \pm 0.006$, and also with the
predictions of supersymmetric grand unification.  It is somewhat higher,
however, then some low energy determinations based on deep inelastic
scattering or lattice calculations of the $c\bar{c}$ or $b\bar{b}$
spectra.  Although the value in (\ref{eqd}) is  reliable
assuming the standard model, it is sensitive to new physics in the $Z \ra
b\bar{b}$ vertex, which is experimentally high.  If new physics contributes to
that vertex there can be a smaller $\alpha_s$, in better agreement with
the low energy determinations.

\item The data exhibits some preference for a light $M_H$, consistent with
the expectations of supersymmetry.  However, this evidence is weak
statistically, with values up to $\sim 800$~GeV still allowed.  Furthermore,
most of the sensitivity to $M_H$ is due to the input values of $R_b$ and
$A_{LR}$, both of which are high compared to the standard model
predictions.  For these reasons, one should not take the $M_H$ limit too
seriously at present.

\item The combination of the precision data with the CDF direct
determination of $m_t$ allows one, for the first time,
to separate the contributions of certain types of new physics from the
dependence of $m_t$.  In particular, the parameters $\rho_0 -1$, $S_{\rm
new}$, $T_{\rm new}$, and $U_{\rm new}$ which describe various types of
$SU_2$ breaking beyond the standard model are all
consistent with zero and stringently constrained~\cite{C}.

\item  Most theories involving compositeness or dynamical symmetry breaking
are strongly disfavored by the precision experiments as well as the
nonobservation of FCNC.

\item In supersymmetry, on the other hand, for most of the allowed
parameter space the heavy particles decouple from FCNC, precision
experiments, and CP violation, consistent with the nonobservation of
deviations.  Another implication is that $M_H < 150$~GeV, and in many cases
the limit is still lower.  This implies that the light Higgs may be
observable in the immediate future at accelerators.  The coupling constant
unification predicts $\alpha_s = 0.129(8)$, where the uncertainty is a
typical (and non-rigorous) range of theoretical uncertainties associated
with the low and high scale thresholds~\cite{pol1}.
The central value is in good
agreement with the LEP determinations from the line shape and jet event
shapes.  However, given the theoretical uncertainties, lower values, as
suggested by low energy data, are possible.  If one has a true
supersymmetric grand unification, then depending on assumptions about the
Higgs sector there may be interesting predictions for the ratio $\tan
\beta$ of the two Higgs vacuum expectation values and for the Higgs mass
$M_H$.  The models often predict an observable proton decay rate, and there
are model dependent but nevertheless interesting predictions for cold dark
matter and for neutrino mass.  The ideas of supersymmetric unification are,
in many ways, suggestive of superstring theories.  However, the precise
connection between the two sets of ideas still escapes us.

\end{itemize}

\end{document}